\documentclass[a4paper,12pt]{article}
\def\R{{\bf R}}
\def\1{\'{\i}}

\begin{document}

\title{\bf Number and amplitude of limit cycles emerging
from {\it topologically equivalent} perturbed centers}
\author{J.L. L\'{o}pez$^{\dag}$ and R. L\'{o}pez-Ruiz$^{\ddag}$\\
                                  \\
{$^{\dag}$\small Department of Mathematics,} \\
{\small Universidad P\'{u}blica de Navarra, 31006-Pamplona (Spain)} \\
{$^{\ddag}$\small Department of Computer Science,} \\
{\small Faculty of Sciences, Universidad de Zaragoza,
50009-Zaragoza (Spain)}
\date{ }}

\maketitle
\baselineskip 8mm
\begin{center} {\bf Abstract} \end{center}
We consider three examples of weekly perturbed centers which
do not have {\it geometrical equivalence}: a linear center,
a degenerate center and a non-hamiltonian center. In each case
the number and amplitude of the limit cycles emerging from
the period annulus are calculated following the same strategy:
we reduce of all of them to locally equivalent perturbed integrable
systems of the form: $dH(x,y)+\epsilon(f(x,y)dy-g(x,y)dx)=0$,
with $H(x,y)=\frac{1}{2}(x^2+y^2)$. This
reduction allows us to find the Melnikov
function, $M(h)=\int_{H=h}fdy-gdx$, associated to each particular
problem. We obtain the information on the bifurcation curves of
the limit cycles by solving explicitly the equation $M(h)=0$
in each case.
$\;$\par\noindent

\newpage
\section{Introduction}

Dissipation is present in almost every physical phenomenon.
The only possibility to create self-sustained oscillations
in a system is by re-injecting the energy dissipated into it.
For instance, the pumping in a switched on laser must
provide the necessary energy to compensate the lost energy
ought to
the interaction between the single frequency light beam
and its surrounding (cavity, mirrors, etc.).
In other context, the solar nuclear bomb must radiate
to the space enough energy to maintain all the living cycles
on the earth. In this case,
many different frequencies are associated to the
thousands of different species of animals and plants \cite{winfree}.
Thus, every oscillation in nature is generated by a
balance between input and output energies, in definitive
between amplification and dissipation.
This dynamical state (characterized by a preferred period,
wave form and amplitude) is represented in phase space
by a isolated closed curve called {\it limit cycle} \cite{andronov}.
This isolated periodic motion can emerge from the perturbation
of a conservative system where a continuum of periodic solutions
exists. Only those closed orbits where the energetic balance
after perturbation vanishes remain as periodic motions.
Take, for instance, the harmonic oscillator $\ddot{x}+x=0$.
A pattern of circles filling the whole plane $(x,\dot{x})$
are all the solutions of this system. A weak nonlinear perturbation
proportional to the velocity, such as that in the Li\'enard equation
$\ddot{x}+\epsilon f(x)\dot{x}+x=0$, destroys the geometrical aspect
of the phase space.
The system can balance the pumping and damping caused
by the nonlinear term, $\epsilon f(x)\dot{x}$,
only on some special periodic orbits.
In the van der Pol case, where $f(x)=(1-x^2)$,
it has been proved the existence, uniqueness and non-algebraicity
of a limit cycle for every value of the control
parameter $\epsilon$ \cite{odani}. Depending on the conditions
imposed over the strength of the nonlinearity $\epsilon$ and
over the properties of $f(x)$: degree, parity, etc.,
many results on the number and amplitude of the limit cycles
for general Li\'enard systems are scattered in the literature
(see, for instance,
Ref. \cite{rychkov,lins,lloyd,giacomini,lopez,lopez1,depassier}
and references therein). This behavior is the mathematical
evidence of how a small perturbation in a physical system
can destroy the coexistence of infinitely many periodic motions
and leaves only a finite number of them verifying strict
energetic balance conditions.

Following this line of reasoning, the aim of this work is to exploit
the topological similarities of different systems presenting an
infinite number of periodic oscillations which give rise to the emergence
of isolated closed orbits under small perturbations.
This is an alternative perspective of the problem of identifying
the limit cycles growing from centers respect to other methods existing
in the literature \cite{gaiko,chicone,blows,giacomini1}.
In Section 2 we sketch the theory behind this type of systems
and we develop the method to find those periodic solutions.
In section 3 we apply this method to three different problems
and calculate the number and amplitude of the limit cycles
emerging from their perturbations. The comparison with previous
investigations on these systems is also performed. Finally,
we present our conclusions.

\section{Geometrical and Topological Equivalence}

The plane is the natural space for the representation of
the integral curves of a second order one-dimensional or
first order two-dimensional dynamical system.
A periodic motion of this system draws a closed orbit on
the plane. If the relation $H(x,y)=h$, with $h$ a fixed real number,
is verified by every point $(x,y)$ on this type of orbit,
a continuum set of periodic orbits is obtained when $h$ runs on
an interval of real values, $h\in[h_1,h_2]$.
This orbit structure receives the name of {\it period annulus}.
Suppose, without lose of generality, that
the boundary orbit given by $H=h_1$ is a degenerate orbit,
say the origin $(0,0)$. This equilibrium point surrounded
in its immediate neighborhood by closed paths is called
a {\it center} \cite{chavarriga}. Obviously, every orbit
of the period annulus verifies the differential relation
\begin{equation}
dH = 0.
\label{eq:dH}
\end{equation}

Depending on the time parameterization of the curves $H(x,y)=h$,
different dynamical systems having the same pattern of integral paths
are obtained. At a first sight,
the difficulty to find the integrating factor
in each case prevent us from establishing the
phase plane geometrical equivalence between them:
we say that two systems are {\it phase plane geometrically
equivalent} when their integral curves on the plane are the same.
Think, for instance, on the set of circles
$H_c(x,y)=\frac{1}{2}(x^2+y^2)=h$.
They obey the differential relation $dH_c=2xdx+2ydy=0$.
Different time parameterizations of this expression
produce systems with different time behaviors
(e.g.: (1) $\ddot{x}+x=0$ ;
(2) $\dot{x}=ky$, $\dot{y}=-kx$, $k=cte.$, and
(3) $\dot{x}=y^2$, $\dot{y}=-xy$),
although with the circles as their integral curves.
If the interest does not reside on the time evolution
of the system, the last concept of geometrical equivalence
will take importance in the study of the system.

One step further. It is also possible that two different
planar systems (having each one a period annulus,
given by $H(x,y)=h$ and $\bar{H}(x,y)=\bar{h}$, with $h\in[h_1,h_2]$ and
$\bar{h}\in[\bar{h}_1,\bar{h}_2]$, respectively) which are
not phase plane geometrical equivalent,
are yet {\it topologically equivalent}. By this we mean that there exists
a bijective and continuous transformation of coordinates (homomorphism),
$\Gamma:\R^2\rightarrow\R^2$, transforming one period annulus
into the other one. That is, if we represent the integral curves
of the first system as $C_h=\{(x,y)\mid H(x,y)=h\}$ and of the second one
as $\bar{C}_{\bar{h}}=\{(x,y)\mid \bar{H}(x,y)=\bar{h}\}$ then
$\Gamma(\bar{C}_{\bar{h}})=C_h$ in a continuous and monotone way
when $h$ and $\bar{h}$ run in their intervals of existence.
In particular, the boundary conditions
over $\Gamma$ are: $\Gamma({\bar{C}_{\bar{h}_1}})=C_{h_1}$ and
$\Gamma({\bar{C}_{\bar{h}_2}})=C_{h_2}$.
Take for example the system $\dot{x}=ay$, $\dot{y}=-bx$ with $a,b>0$.
It presents a period annulus formed by ellipses,
$H(x,y)=\frac{x^2}{a}+\frac{y^2}{b}=h$. Clearly, this system is
topologically equivalent to another one having a circular period annulus:
$\bar{H}_c(x,y)=\frac{1}{2}(x^2+y^2)=\bar{h}$.
It is straightforward to verify that the function
$\Gamma(x,y)=(\frac{x}{\sqrt{a}},\frac{y}{\sqrt{b}})$
establishes the homomorphism between both systems.
Obviously, if two systems are
geometrical equivalent they are topologically equivalent because in this case
$H=\bar{H}$ and then we can choose $\Gamma$ as the identity mapping.

The relevance of this property is the following. Imagine that we are
able to find the limit cycles that emerge
from the perturbation of the first system,
\begin{equation}
dH + \hbox{perturbation} = 0.
\label{eq:dHp}
\end{equation}
Then, the application of $\Gamma$ to the second perturbed system,
$d\bar{H} + \hbox{perturbation'} = 0$, bring this last
expression to the form (\ref{eq:dHp}), even if we do not know $\bar{H}$.
Now the limit cycles of the second system can be found
by solving equation (\ref{eq:dHp}). Undoing the change of variables
with $\Gamma^{-1}$, we will obtain the periodic motions of the
perturbed system in the original coordinates.

\section{Limit Cycles emerging from some Centers}

\subsection{Perturbation of a circular period annulus}

As an application of the method above explained, we study
here the limit cycles emerging from the perturbation of what
we could consider a paradigmatic case in the center's typology:
a continuum set of circles,
$H_c(x,y)=\frac{1}{2}(x^2+y^2)=h$.
The solution of many different dynamical systems is this
kind of period annulus. As a simple representation of
them we take the linear center:
\begin{equation}
\begin{array}{l}
\dot{x}=y, \\
\dot{y}=-x,
\end{array}
\label{eq:center00}
\end{equation}
whose integral curves clearly verify $dH_c=x\dot{x}+y\dot{y}=0$.
This path diagram is destroyed when we slightly perturb
this conservative system with two general nonlinear terms,
$f(x,y)$ and $g(x,y)$, controlled by the
parameter $\mid\epsilon\mid\ll 1$,
\begin{equation}
\begin{array}{l}
\dot{x}=y+\epsilon f(x,y), \\
\dot{y}=-x+\epsilon g(x,y).
\end{array}
\label{eq:center000}
\end{equation}
It is a time parameterized version of the differential relation:
$dH_c+\epsilon\,(f\,dy-g\,dx)=0$, in which the time variable has
been removed.
Following the procedure proposed in the previous section,
we need to study this differential relation in order to find
the limit cycles of system (\ref{eq:center000}).
If its integral curves are represented by $y(x)$ and we define
$y'(x)=dy/dx$, we obtain:
\begin{equation}
(yy'+x)+\epsilon [f(x,y)y'-g(x,y)]= 0.
\label{eq:center0}
\end{equation}
If we suppose the origin $(0,0)$ is the only fixed
point of equation (\ref{eq:center000}),
a limit cycle $C_l\equiv (x,y_{\pm}(x))$ of it,
with a positive branch $y_+(x)>0$ and a negative branch $y_-(x)<0$,
cut the $x$-axis in two points $(-a_-,0)$ and $(a_+,0)$
with $a_-,a_+>0$. Every limit cycle $C_l$ solution
of equation (\ref{eq:center000}) encloses the origin and
the oscillation $x$ runs in the interval $-a_-<x<a_+$.

The amplitudes of oscillation $a_-,a_+$ identify the limit cycle.
The result for $\epsilon\neq 0$ is a nested set of closed curves
that defines the qualitative distribution
of the integral curves in the plane $(x,y)$. The stability
of the limit cycles is alternated. For a given stable limit cycle,
the two neighboring limit cycles, the closest one in its interior
and the closest one in its exterior, are unstable,
and conversely (see Figure 1). If we determine the stability
of the origin, then the stability of every limit cycle
remains fixed by the alternation property.
The stability of the origin is determined
by the sign of the real part of
the eigenvalues of the Jacobian at this point.
Hence, its calculation shows that if
$\epsilon(f_x(0,0)+g_y(0,0))<0$ then the origin is stable
and if $\epsilon(f_x(0,0)+g_y(0,0))>0$ then it is unstable.

Another property of a limit cycle can be derived from
the fact that the mechanical energy
$E=H_c=\frac{1}{2}(x^2+y^2)/2$ is conserved in a whole oscillation:
\begin{displaymath}
\int_{C_l}dH_c=\int_{C_l}\frac{dE}{dx} dx = 0.
\end{displaymath}
Thus, if equation (\ref{eq:center0}) is integrated
along a limit cycle, between the amplitudes of oscillation,
we obtain:
\begin{equation}
\begin{array}{cl}
\int_{a_-}^{a_+}[g(x,y_+(x))-f(x,y_+(x))y'_+(x)] dx \;+\; \\
\int_{a_+}^{a_-}[g(x,y_-(x))-f(x,y_-(x))y'_-(x)] dx= 0.
\end{array}
\label{eq:ciclos0}
\end{equation}
Every couple of solutions $\lbrace y_+(x),y_-(x)\rbrace$ of
equation (\ref{eq:ciclos0}), vanishing in the extremes,
$y_+(a_-)=y_-(a_-)=y_+(a_+)=y_-(a_+)=0$,
and verifying equation (\ref{eq:center0}), constitute
the finite set of limit cycles of equation (\ref{eq:center000}).
As a consequence of the continuity in the parameter $\epsilon$,
these conditions are also valid for $\epsilon=0$.
For this parameter value, the limit cycles are circles:
$y_+(x)=\sqrt{a^2-x^2}$ and $y_-(x)=-\sqrt{a^2-x^2}$.
Here $a_+=a_-=a>0$ represents the amplitudes of the limit cycles.
At this order of approximation, the condition (\ref{eq:center0})
is verified, and condition (\ref{eq:ciclos0}) reads:
\begin{equation}
\begin{array}{c}
\int_{-a}^{a}\left[g(x,\sqrt{a^2-x^2})+
{xf(x,\sqrt{a^2-x^2})\over\sqrt{a^2-x^2}}\right] dx + \\
\int_{a}^{-a}\left[g(x,-\sqrt{a^2-x^2})-
{xf(x,-\sqrt{a^2-x^2})\over\sqrt{a^2-x^2}}\right] dx= 0.
\end{array}
\label{eq:beta00}
\end{equation}
that is,
\begin{equation}
\beta(a)\equiv \int_{-a}^{a}\left[\tilde g(x,\sqrt{a^2-x^2})+
x{\tilde f(x,\sqrt{a^2-x^2})\over\sqrt{a^2-x^2}}\right] dx = 0,
\label{eq:beta0}
\end{equation}
where
\begin{equation}
\begin{array}{c}
\tilde f(x,y)\equiv {1\over 2}[f(x,y)+f(x,-y)-f(-x,y)-f(-x,-y)],\\
\tilde g(x,y)\equiv {1\over 2}[g(x,y)-g(x,-y)+g(-x,y)-g(-x,-y)].
\end{array}
\label{eq:fg0}
\end{equation}
If $\beta(a)$ does not vanish identically,
each solution $a>0$ of the equation $\beta(a)=0$ is the amplitude
of a limit cycle of the system (\ref{eq:center000}) in the weak nonlinear
regime. And conversely, at order zero in $\epsilon$,
the amplitudes of all the limit cycles emerging from the period
annulus are solutions of equation (\ref{eq:beta0}).
These results are exact for $\epsilon =0$. Therefore,
equations (\ref{eq:beta00}) or (\ref{eq:beta0}) determine
the amplitudes of the period motions surviving to a slight perturbation
of a period annulus formed by circles. A typical portrait of limit cycles in
this regime is given in Figure 1.

We remark that $\beta(a)$ is the first Melnikov function of
the system (\ref{eq:center000}).
Here it has been obtained by following a different approach
to the usual one (based on the calculation of the displacement function
of the first return mapping on the period annulus).
Observe also that $\tilde f(x,y)$ is an odd function
of $x$ and an even function of $y$, whereas $\tilde g(x,y)$
is an even function of $x$ and an odd function of $y$.
Therefore, if $f(x,y)$ and $g(x,y)$ are polynomials in $x$ and $y$,
only the odd terms in $x$ and even in $y$ of $f(x,y)$
and the even terms in $x$ and odd in $y$ of $g(x,y)$
survive in (\ref{eq:fg0}) and contribute to $\beta(a)$
in (\ref{eq:beta0}).

As an example, we integrate equation (\ref{eq:beta0}) when
$f(x,y)$ is a polynomial of degree $2n_1$ or $2n_1+1$ in $x$
and degree $2m_1$ or $2m_1+1$ in $y$
and $g(x,y)$ is a polynomial of degree $2n_2$ or $2n_2+1$ in $x$
and degree $2m_2$ or $2m_2+1$ in $y$. Then,
\begin{displaymath}
\begin{array}{c}
\tilde f(x,y)=\sum_{j=0}^{n_1}\sum_{k=0}^{m_1}a_{j,k}x^{2j+1}y^{2k},\\
\tilde g(x,y)=\sum_{j=0}^{n_2}\sum_{k=0}^{m_2}b_{j,k}x^{2j}y^{2k+1}.
\end{array}
\end{displaymath}
where $a_{j,k}$ and $b_{j,k}$ are real coefficients. The result is
\begin{displaymath}
\beta(a)=\frac{a^2}{2}\sum_{j=0}^{n}\sum_{k=0}^{m}\frac{\Gamma(j+1/2)\Gamma(
k+1/2)}
{(k+j+1)!}\left[\left(j+\frac{1}{2}\right)a_{j,k}+
\left(k+\frac{1}{2}\right)b_{j,k}\right]a^{2(k+j)},
\end{displaymath}
where $n\equiv$Max$\lbrace n_1, n_2\rbrace$ and
$m\equiv$Max$\lbrace m_1, m_2\rbrace$. Here,
$a_{j,k}=0$ for $j>n_1$ or $k>m_1$ and
$b_{j,k}=0$ for $j>n_2$ or $k>m_2$.
The root $a=0$ of $\beta(a)$ corresponds to the fixed point $(0,0)$
and the factor $a^2$ can be eliminated. Thus, the possible
amplitudes $a$ are the zeros of $\beta(a)/a^2$, a polynomial in $a^2$
of degree $Max\lbrace n_1+m_1,n_2+m_2\rbrace$.
There are no more than $Max\lbrace n_1+m_1,n_2+m_2\rbrace$ different solutions
$a>0$ and therefore, the maximum number of limit cycles in this case
is Max$\lbrace n_1+m_1,n_2+m_2\rbrace$ as it was proved by Iliev \cite{iliev}.
This result can also be written as
$\left[\frac{Max\lbrace deg\, f,\, deg\, g\rbrace-1}{2}\right]$
where $[\;]$ means the integer part and $deg\,f$ (resp. $deg\,g$) denotes
the degree of $f(x,y)$ (resp. of $g(x,y)$).
If we restrict ourselves to the case of Li\'enard equations where
$f(x,y)=0$ and $g(x,y)$ is linear in $y$, we recover the
known result that the number of limit cycles is less or equal than $n_2$
in the weak nonlinear regime. The extension of this relation
for the whole range of $\epsilon$ is the still not proved
Lins-Melo-Pugh conjecture on the number of limit cycles
for Li\'enard systems \cite{lins}.

\subsection{Reduction of some examples to a perturbed circular period annulus}

{\bf Example 1: (a particular circular period annulus)}
\begin{equation}
\begin{array}{l}
\dot{x}=y-\epsilon(a_1x+a_2x^2+a_3x^3), \\
\dot{y}=-x-\epsilon b_3x^3,
\end{array}
\label{eq:system1}
\end{equation}

Yuquan and Zhujun \cite{yuquan} have investigated the number
of limit cycles of a similar and more general system.
The conditions of their theorems can be translated for this
particular case of weak perturbations ($\mid\epsilon\mid\ll 1$)
as following: (i) If $\epsilon b_3>0$, $\epsilon a_1<0$
and $\epsilon a_3>0$ then system (\ref{eq:system1})
has a unique stable limit cycle ({\it Theorem 1} in Ref. \cite{yuquan});
(ii) If $\epsilon b_3>0$, $\epsilon a_1>0$,
and $\epsilon a_3<0$ then system (\ref{eq:system1})
has a unique unstable limit cycle ({\it Theorem 2} in Ref. \cite{yuquan}).

These results are confirmed by the method introduced in the
previous section. Moreover, as it is explained in that section,
the amplitudes $a$ of the limit cycles in the weakly nonlinear
regime are the nontrivial positive solutions of the equation
$\beta (a)=0$. In this case, $f(x,y)=-a_1x-a_2x^2-a_3x^3$ and
$g(x,y)=-b_3x^3$. We obtain:
\begin{displaymath}
\beta(a)=\frac{\pi}{2}\,a^2(a_1+\frac{3}{4}\,a_3a^2),
\end{displaymath}
which shows that the system has at most one limit cycle
of amplitude
\begin{displaymath}
a=\sqrt{-\frac{4a_1}{3a_3}}\;.
\end{displaymath}
Therefore, the system has not limit cycles
if $a_1a_3> 0$ and just one limit cycle if $a_1a_3<0$.
This periodic motion is stable if $\epsilon a_1<0$
and unstable if $\epsilon a_1>0$.
Numerical simulations are in strong agreement with this
analytical result.
Observe that there are certain conditions of Yuquan and Zhujun
theorems that can be removed in this regime in order to have
just one limit cycle.

{\bf Example 2: (a system topologically equivalent to a circular period
annulus)}
\begin{equation}
\begin{array}{l}
\dot{x}=-y^{2l-1}+\epsilon P(x,y), \\
\dot{y}=x^{2k-1}+\epsilon Q(x,y),
\end{array}
\label{eq:system2}
\end{equation}

\noindent where $k$ and $l$ are positive integers and
$P(x,y)$ and $Q(x,y)$ are polynomials in $x$ and $y$.
The number of limit cycles of this system for $\mid\epsilon\mid\ll 1$
has been studied in Ref. \cite{llibre,gasull}.
The constant of motion for the unperturbed problem is
$\bar{H}(x,y)=\frac{1}{2k}\,x^{2k}+\frac{1}{2l}\,y^{2l}$.
When $k,l>1$ it corresponds to a degenerate center
topologically equivalent to the circular period annulus
obtained when $k=l=1$ (then $H_c=\frac{1}{2}\,(x^2+y^2)$).
If we remove the time variable, we obtain
\begin{equation}
y^{2l-1}\frac{dy}{dx}+x^{2k-1}+\epsilon \left[Q(x,y)-
P(x,y)\frac{dy}{dx}\right]=0.
\label{eq:system22}
\end{equation}

In order to apply the method introduced in the previous section,
we perform the change of variables $\Gamma: (x,y)\rightarrow (X,Y)$:
\begin{displaymath}
\begin{array}{c}
X=sign(x)\frac{\mid x\mid^k}{\sqrt{k}}, \\
Y=sign(y)\frac{\mid y\mid^l}{\sqrt{l}}.
\end{array}
\end{displaymath}
Then, equation (\ref{eq:system22}) reads
\begin{equation}
Y\frac{dY}{dX}+X+\epsilon \left[\frac{\mid X\mid^{(1/k)-1}}
{k^{1-(1/2k)}}
Q\left(x,y\right)-
\right. \\ \left.
\frac{\mid Y\mid^{(1/l)-1}}{l^{1-(1/2l)}}
P\left(x,y\right)
\frac{dY}{dX}\right]=0,
\label{eq:vv}
\end{equation}
with
\begin{equation}
\begin{array}{c}
x=sign(X)(\sqrt{k}\mid X\mid)^{1/k}
\\
y=sign(Y)(\sqrt{l}\mid Y\mid)^{1/l}.
\end{array}
\end{equation}
Equation (\ref{eq:vv}) has the form (\ref{eq:center0})
with $(x,y)$ replaced by $(X,Y)$ and
\begin{equation}
\begin{array}{c}
f(X,Y)=-\frac{\mid Y\mid^{(1/l)-1}}{1-l^{(1/2l)}}
P\left(sign(X)(\sqrt{k}\mid X\mid)^{1/k},
sign(Y)(\sqrt{l}\mid Y\mid)^{1/l}\right) \\
g(X,Y)=-\frac{\mid X\mid^{(1/k)-1}}{1-k^{(1/2k)}}
Q\left(sign(X)(\sqrt{k}\mid X\mid)^{1/k},
sign(Y)(\sqrt{l}\mid Y\mid)^{1/l}\right).
\end{array}
\label{eq:fg1}
\end{equation}
Therefore, in these coordinates,
if $A$ is the amplitude of a limit cycle
$(X,Y(X))$ in the weakly nonlinear regime,
it verifies the equation $\bar\beta(A)=0$, where
\begin{displaymath}
\bar\beta(A)\equiv\int_{-A}^A \left\lbrace \tilde
g\left(X,Y_A(X)\right)+ \frac{X}{Y_A(X)}\tilde
f\left(X,Y_A(X)\right)\right\rbrace dX,
\end{displaymath}
$\tilde f\left(X,Y\right)$ and $\tilde g\left(X,Y\right)$
are defined in (\ref{eq:fg0}) with $f\left(X,Y\right)$
and $g\left(X,Y\right)$ given in (\ref{eq:fg1}) and
$Y_A(X)\equiv\sqrt{A^2-X^2}$. If we undo the change of variable
$(X,Y)\to(x,y)$ in the above integral and write $a^{2k}\equiv kA^2$, then
we can write $\bar\beta(A)\equiv\beta(a)$ in terms of the amplitudes $a$ of
the original variables $(x,y)$:
\begin{displaymath}
\beta(a)\equiv\int_{-a}^a \left\lbrace \tilde
Q\left(x,y_a(x)\right)-\tilde
P\left(x,y_a(x)\right)\frac{dy_a(x)}{dx}\right\rbrace dx,
\end{displaymath}
where $y_a(x)\equiv\left[\frac{l}{k}\left(a^{2k}-x^{2k}\right)\right]^{1/(2l)}$
and $\tilde P(x,y)$ and $\tilde Q(x,y)$ are defined in (\ref{eq:fg0}) for
$f(x,y)=P(x,y)$ and $g(x,y)=Q(x,y)$.
Hence, the amplitudes, $a$, of the limit cycles in the weakly nonlinear regime
are the nontrivial solutions of the equation $\beta (a)=0$.

If $k=l$, then $\beta(a)$ reads
\begin{displaymath}
\beta(a)=a\int_{-1}^1 \left\lbrace \tilde
Q\left(ax,a\tilde y(x)\right)+ \frac{x^{2k-1}\tilde y(x)\tilde
P\left(ax,a\tilde y(x)\right)}{1-x^{2k}}\right\rbrace dx,
\end{displaymath}
where $\tilde y(x)\equiv\left(1-x^{2k}\right)^{1/(2k)}$.
If $P(x,y)$ and $Q(x,y)$ are polynomials of degree at most $n$,
$a^{-2}\beta(a)$ is a polynomial of degree
$\lfloor\frac{n-1}{2}\rfloor$ in $a^2$
and we conclude that the maximum number of limit cycles in this regime is
$\lfloor\frac{n-1}{2}\rfloor$, as
it has been shown by Coll {et al.} \cite{gasull}.

Take, for instance, the particular case:
\begin{equation}
\begin{array}{c}
\dot{x} =  -y^3+\epsilon bxy^2, \\
\dot{y} =  x^3+\epsilon cx^2y^3.
\end{array}
\label{eq:system222}
\end{equation}
with  $k=l=2$ and $n=5$.
The equation $\beta(a)=0$ has at most one non trivial solution given by
\begin{displaymath}
a=\,\frac{(\pi/2)^{3/4}}{\Gamma(3/4)}\;\sqrt{\frac{-b}{c}},
\end{displaymath}
In this case, the system has at most one limit cycle
of amplitude $a$ emerging from the slightly perturbed period annulus.
This analytical result is in agreement with the direct verification
of the dynamics by integrating the equations (\ref{eq:system222}).

{\bf Example 3: (a system topologically equivalent to a circular period
annulus)}
\begin{equation}
\begin{array}{l}
\dot{x}=-y+yx^2+\epsilon F(x,y), \\
\dot{y}=x+xy^2+\epsilon G(x,y),
\end{array}
\label{eq:system3}
\end{equation}

\noindent where $F(x,y)=a_0x^3+a_1x^2y+a_2xy^2+a_3y^3$
and $G(x,y)=b_0x^3+b_1x^2y+b_2xy^2+b_3y^3$.
It has been shown that this system has, at most, two
limit cycles ({\it theorem 2.6} in Ref. \cite{llibre1}).\newline
\indent The constant of motion of equation (\ref{eq:system3})
is $H(x,y)=\frac{1+y^2}{1-x^2}=h$.
A period annulus topologically equivalent to
$\bar{H}_c=\frac{1}{2}\,(x^2+y^2)=\bar{h}$
is defined in the region $-1<x<1$ when $h$ runs from $1$ to $\infty$.
The origin $(0,0)$ is found for $h=1$ and the boundaries of this region
are the lines $x=\pm 1$ which are approached when $h\rightarrow\infty$.
The interior is formed by closed curves topologically equivalent to circles.
The integral paths exterior to the period annulus are open curves
obtained when $h$ increases from $-\infty$ to zero.

Following the same steps as in the preceding example,
we proceed to remove the time variable of system (\ref{eq:system3}).
Writing $x'\equiv\frac{dx}{dy}$:
\begin{equation}
(1+y^2)xx'+(1-x^2)y+\epsilon[G(x,y)x'-F(x,y)]=0.
\label{eq:system33}
\end{equation}
The homomorphism with the circular period annulus is established
by means of the change of variables $\Gamma: (x,y)\rightarrow (X,Y)$:
\begin{displaymath}
\begin{array}{cl}
\;\;X=Sign(x)\sqrt{-\log(1-x^2)}, \\
Y=Sign(y)\sqrt{\log(1+y^2)}.
\end{array}
\end{displaymath}
Then, the equation (\ref{eq:system33}) reads
\begin{displaymath}
\begin{array}{cl}
X\frac{dX}{dY}+Y+\epsilon \left[\mid X\mid\frac{dX}{dY}\frac{
G\left(Sign(X)\sqrt{1-e^{-X^2}},Sign(Y)\sqrt{e^{Y^2}-1}\right)}
{e^{Y^2}\sqrt{1-e^{-X^2}}}-
\right. \\ \left.
\frac{\mid Y\mid F\left(Sign(X)\sqrt{1-e^{-X^2}},Sign(Y)
\sqrt{e^{Y^2}-1}\right)}
{e^{-X^2}\sqrt{e^{Y^2}-1}}\right]=0.
\end{array}
\end{displaymath}
Now this equation has the form (\ref{eq:center0})
with the variables $(x,y)$ replaced by the variables $(Y,X)$
and
\begin{equation}
\begin{array}{c}
f(Y,X) = \mid X\mid
\frac{G\left(Sign(X)\sqrt{1-e^{-X^2}},Sign(Y)\sqrt{e^{Y^2}-1}\right)}
{e^{Y^2}\sqrt{1-e^{-X^2}}}, \\
g(Y,X) = \mid Y\mid \frac{F\left(Sign(X)\sqrt{1-e^{-X^2}},Sign(Y)
\sqrt{e^{Y^2}-1}\right)}
{e^{-X^2}\sqrt{e^{Y^2}-1}}.
\end{array}
\end{equation}
Therefore,
we can apply the method of the previous section to this equation.
Then, the amplitude $A$ of a limit cycle $(Y,X(Y))$ of this system
in the weakly nonlinear regime,
$X(Y)\simeq\sqrt{A^2-Y^2}$, satisfies $\bar\beta(A)=0$, with
\begin{displaymath}
\begin{array}{c}
\bar\beta(A)\equiv\int_{-A}^A \left\lbrace
\frac{\mid Y\mid\tilde F\left(\sqrt{1-e^{Y^2-A^2}},\sqrt{e^{Y^2}-1}\right)}
{e^{Y^2-A^2}\sqrt{e^{Y^2}-1}}+
\frac{Y\tilde G\left(\sqrt{1-e^{Y^2-A^2}},\sqrt{e^{Y^2}-1}\right)}
{e^{Y^2}\sqrt{1-e^{Y^2-A^2}}}
\right\rbrace dY ,
\end{array}
\end{displaymath}
$\tilde F(x,y)\equiv a_0x^3+a_2xy^2$
and $\tilde G(x,y)\equiv b_1x^2y+b_3y^3$.
If we undo the change of variable
$(Y,X)\to(x,y)$ in the above integral and write $A^2=\log(1+a^2)$, then
we can write $\bar\beta(A)\equiv\beta(a)$ in terms of the amplitudes $a$ of
the original variables $(x,y)$:
\begin{displaymath}
\beta(a)\equiv\int_{-a}^a \left\lbrace
\frac{y\sqrt{1+a^2}\,\tilde G\left(\sqrt{\frac{a^2-y^2}{1+a^2}},y\right)}
{\sqrt{a^2-y^2}}+(1+a^2)\tilde F\left(\sqrt{\frac{a^2-y^2}{1+a^2}},y\right)
\right\rbrace \frac{dy}{(1+y^2)^2}.
\end{displaymath}
After straightforward computations we obtain that the
real roots of $\beta(a)$ are the real solutions of the
equation $\tilde\beta(a)=0$, where
\begin{displaymath}
\begin{array}{c}
\tilde\beta(a)\equiv [b_1+a_2(1+a^2)]\,[2+a^2-2\sqrt{1+a^2}]\;+ \\ \\
b_3[2(1+a^2)^{3/2}-3a^2-2] + a_0[2\sqrt{1+a^2}-2-a^2+a^4].
\end{array}
\end{displaymath}
We replace the variable $a$ in this equation by the variable
$B\equiv\sqrt{1+a^2}$. We look for solutions
$a>0$, that is, $B>1$. The two only possible solutions are
\begin{displaymath}
B^\pm\equiv\frac{-(a_0+b_3)\pm\sqrt{(a_0+b_3)^2-(a_0+a_2)(b_1+b_3)}}{a_0+a_2}.
\end{displaymath}
Writing
\begin{equation}
\begin{array}{c}
u\equiv\frac{a_0+b_3}{a_0+a_2}, \;\;\;\;
v\equiv\frac{b_1+b_3}{a_0+a_2},
\end{array}
\end{equation}
we have that $B^\pm=-u\pm\sqrt{u^2-v}$. Then, the number of limit cycles
in the weakly nonlinear regime is $0$, $1$ or $2$ depending on
$a_0,a_2,b_1,b_3$ (see figure 2):

(i) If $u>-1$ and $v+2u+1>0$ or $v>u^2$, neither $B^+$ neither $B^-$
are greater than $1$. Therefore,
the system has not limit cycles.\newline
\indent (ii) If $v+2u+1<0$, then $B^+>1$, $B^-<1$. Therefore,
the system has a unique limit cycle with amplitude $a=\sqrt{(B^+)^2-1}$.\newline
\indent (iii) If $v+2u+1>0$ and $v<u^2$, then $B^+,B^->1$. Therefore,
the system has two limit cycles with amplitudes $a^+=\sqrt{(B^+)^2-1}$
and $a^-=\sqrt{(B^-)^2-1}$.

Computer integration of equation (\ref{eq:system3}) is
in agreement with these analytical results.
Theorem 2.6 of Ref. \cite{llibre1} states that the maximum
number of limit cycles of this system is two. By applying
the method proposed in Section 2 we obtain a more detailed
information: the exact number
and amplitudes of the limit cycles of this system as a function
of the parameters $\{a_k, b_k\mid k=1,2,3\}$.

\section{Conclusions}

\indent Nowadays it is well known that the inclusion of a slight dissipation
in a conservative system destroys almost completely the geometrical aspect
of the phase space of the unperturbed motion.
But the analytical methods available for the study
of such perturbed systems suffer a lack of predictive power
when the nonlinearities start to dominate the dynamics.

In this work, we have carried out the calculation of the
precise number and amplitude of the periodic orbits surviving
to a weak perturbation of different topologically equivalent
period annulus.
First, the equation $\beta(a)=0$ containing that information
about the limit cycles emerging from a general slight
perturbation of a continuum set of circles is stated.
Essentially this equation corresponds to the first
Melnikov function associated to the system,
although it has been obtained by an alternative line of reasoning.
Second, the dynamical equations of other topologically equivalent
perturbed centers have been reduced to equivalent
equations of the former case.
Then, the calculation of the number and amplitude
of their limit cycles has also been carried out in the same way.
The numerical integration of these different systems
supports the analytical calculations
and confirms the theoretical predictions.

We must stress that, on the one hand, our method has given
the correct maximum number of limit cycles in all the examples
analyzed. On the other hand, it has allowed us to
predict the exact number and amplitude of the limit cycles
emerging from different topologically equivalent perturbed period annulus.
It seems that in order to find the number and amplitude
of limit cycles, it is not important to maintain the condition
of a planar system to be a polynomial system. Perhaps it would be
more interesting to try to identify the class of all its
geometrical and  topologically equivalent systems and
to perform the calculations in the simplest system of this class.

{\bf Acknowledgements:}
J.L. L\'opez acknowledges DGCYT (BFM2000-0803) and Gobierno de Navarra
(Res. 92/2002) for financial support.

\newpage
\begin{center} {\bf Figure Captions} \end{center}

{\bf Figure 1}: A typical phase portrait of equation (\ref{eq:center0}).
The limit cycles of amplitudes $(-a_1,a_2)$, $(-b_1,b_2)$, $\dots$,
enclose the origin. Stable and unstable limit cycles alternate.
The order of this alternation depends on the stability of the origin.

{\bf Figure 2}: The complete bifurcation diagram
of system (\ref{eq:system3}). The system has no
periodic solutions in the white region, one limit cycle
in the grey region and two limit cycles in the black region.
See example 3 for details of the calculation of $(u,v)$
from the original parameters of equation (\ref{eq:system3}).

\end{document}